# Coconut Libtool: Bridging Textual Analysis Gaps for Non-Programmers

| | |
|---|---|
| **Santosa, Faizhal Arif** | National Research and Innovation Agency, Indonesia \| faiz005@brin.go.id |
| **Lamba, Manika** | University of Illinois Urbana-Champaign, USA \| manika@illinois.edu |
| **George, Crissandra** | Case Western Reserve University, USA \| cxg570@case.edu |
| **Downie, J. Stephen** | University of Illinois Urbana-Champaign, USA \| jdownie@illinois.edu |

## ABSTRACT

In the era of big and ubiquitous data, professionals and students alike are finding themselves needing to perform a number of textual analysis tasks. Historically, the general lack of statistical expertise and programming skills has stopped many with humanities or social sciences backgrounds from performing and fully benefiting from such analyses. Thus, we introduce Coconut Libtool (www.coconut-libtool.com/), an open-source, web-based application that utilizes state-of-the-art natural language processing (NLP) technologies. Coconut Libtool analyzes text data from customized files and bibliographic databases such as Web of Science, Scopus, and Lens. Users can verify which functions can be performed with the data they have. Coconut Libtool deploys multiple algorithmic NLP techniques at the backend, including topic modeling (LDA, Biterm, and BERTopic algorithms), network graph visualization, keyword lemmatization, and sunburst visualization. Coconut Libtool is the people-first web application designed to be used by professionals, researchers, and students in the information sciences, digital humanities, and computational social sciences domains to promote transparency, reproducibility, accessibility, reciprocity, and responsibility in research practices.

## INTRODUCTION

Across almost every information organization is a rush of discussion surrounding artificial intelligence (AI), large language models (LLMs), natural language processing (NLP), and more, demanding that information professionals adapt to these changes, lead discussions, and contribute to these technological innovations. Within this surge of advancement, many textual analysis tools have been created that integrate AI, LLMs, NLP, and machine learning algorithms, combining both data analysis and visualization within one package. This integration of both textual analysis and visualization results in "even more effective analysis of text data by making it easier to identify patterns, outliers, and correlations that may not be readily apparent from the raw text alone" (Milev, 2023). TechMiner (Velasquez, 2023), PyblioNet (Müller, 2023), Pybliometrics (Rose & Kitchin, 2019), pyBibX (Pereira et al., 2023), metaknowledge (McLevey & McIlroy-Young, 2017), and litstudy (Heldens et al., 2022) are a few examples of tools that have been developed to add value to the analysis of textual data. Other tools that can analyze text from various sources include MoreThanSentiments (Jiang & Srinivasan, 2023) and TAll (Aria et al., 2023). It is necessary to have a basic understanding of coding in order to use litstudy, metaknowledge, MoreThanSentiments, and pyBibX. Furthermore, some must be installed before doing different analyses, which makes having adequate computational resources essential.

With the advancements in research tools, an identified issue that has been highly prevalent is the barriers to access. Open science has pushed to minimize these barriers by creating and promoting openly available tools, scholarly publications, datasets, and more, resulting in a larger transition and transformation of open access across the library and information science field (ACRL, 2018; ACRL, 2020). This transition and transformation have pushed the usage of open-source software, technology, and techniques for textual analysis with the goal of fostering greater transparency, collaboration, accessibility, reciprocity, and responsibility. R, Python, MALLET, KH Coder, RapidMiner, WEKA, Orange, and Voyant Tools are a few examples of popular open-source tools available for researchers, with R and various R-packages being widely well-known due to the high popularity and availability of this analysis and visualization tool (Thakur & Kumar, 2021). R-packages have even been further developed for enhanced automated analyses, such as the open-source bibliometrix R-package (Aria & Cuccurullo, 2017), adding flexibility and swift integration for better user engagement. Open-source textual analysis tools with a simple graphical user interface, like R and others mentioned, serve as a great solution for users to begin implementing these techniques into their daily work and research (Lamba & Madhusudhan, 2022). With data mining and textual analysis being widely adopted, resulting in an explosion of information, these difficult barriers of access still demonstrate use cases that demand further improvement in a variety of sectors, including "differences in library types, financial constraints, technological variations, human resource disparities, and varying levels of staff experience" (Santosa, Lamba, George, 2023). These powerful tools are only effective when users know how to use them (Seadle, 2016), which is where limited competency in computing and programming plays a huge role in these barriers. In a systematic review of data mining and textual analysis scholarly publications, Thakur and Kumar (2021) uncovered multiple barriers to access to these tools, such as a "lack of skills, limited data sets in some domains, legal insecurity, technical problems, improper infrastructure and support for conducting text mining on large corpus size." While all these barriers cannot be solved with one tool or one improvement, it demonstrates the need for continued innovation across the field of data mining and textual analysis by addressing these highly prevalent barriers to further enhance the accessibility of these timely data science methodologies and tools.

In this paper, we aim to address these challenges by introducing Coconut Libtool (www.coconut-libtool.com/), a people-first web application designed to democratize textual analysis with its user-friendly interface. By removing the need for software installation, incorporating a straightforward web-based design, and eliminating the need for coding experience, this open-source tool breaks down the exclusivity surrounding data mining and textual analysis, therefore expanding access to an even wider range of users. First, we will showcase the design of this tool, highlighting the integration of AI, LLMs, NLP, and machine learning algorithms combined with the purposeful integration of user-focused design choices. Along with the design of this tool, we will discuss how the Coconut Libtool addresses challenges currently seen in the research landscape, exemplifying the implications of this open-source, user-friendly, and comprehensive solution that is not only limited to the audience of information professionals but anyone and everyone interested in data mining and textual analysis. It combines some freely accessible textual analysis technologies, including (i) keyword stemming, (ii) bidirectional network analysis, (iii) topic modeling, and to perform (iv) sunburst visualization (Figure 1). The code repository for the Coconut Libtool is available at github.com/faizhalas/library-tools.

## BACKEND FEATURES AND FUNCTIONS

### Data Import and Pre-Processing

Coconut Libtool supports CSV or TXT files sourced from three bibliometric databases: (i) Scopus, (ii) Web of Science, and (iii) Lens.org. Users can also upload their own customized data for textual analysis or visualization. In the process of identifying the suitable type of analysis, particularly for customized files, the app has a menu called Coconut Libtool File Checker. This feature helps to evaluate individual columns within the submitted dataset to facilitate the selection of the desired technique. The data file undergoes a meticulous examination to ensure its completeness prior to parsing and conversion into a structured pandas dataframe for subsequent analysis. Specifically, when importing data files from Web of Science, which often include abbreviations such as 'TI' for 'Title' and 'SO' for 'Source Title', necessary modifications are made. While files sourced from databases like Scopus and Web of Science typically include columns for title, keywords, and abstract, additional fields such as publication year, citation count, document type, and source title are imperative for sunburst visualization. For determining root keywords, NLTK's WordNet Lemmatizer and Snowball Stemmer, accessible to all, serve as valuable tools (Bird, Loper, & Klein, 2009). Furthermore, to cater to user preferences regarding the inclusion of various similar columns such as "Author Keywords" and "Keywords Plus," the 'Keywords Stems' functionality adeptly identifies columns containing the term "keyword."

### Text Analysis

In Coconut Libtool, 'keywords stem' and 'topic modeling' are two of the major text analysis techniques that users can use. 'Keywords stem' is a text pre-processing feature in the app that includes lemmatization & stemming for locating basic words and aids in catching the true meaning of the word, which can lead to improved semantic analysis and comprehension of the text. The app uses NLTK WordNet Lemmatizer to lemmatize sentences and NLTK Snowball Stemmer to stem the sentences. Users can utilize the 'Keyword stem' feature to identify root words. For example, 'apples' and 'apple' are the same, therefore 'apples'will become 'apple.'

Another interesting feature in the Coconut Libtool for text analysis is 'topic modeling'. Users can choose from three different algorithms – (i) LDA (Blei et al., 2003), (ii) Biterm topic model (Yan, Guo, Lan, & Cheng, 2013), and (iii) BERTopic (Grootendorst, 2022) for topic modeling. LDAvis, which is useful for exploring topic-term relationships and is based on the Latent Dirichlet Allocation Model (Sievert & Shirley, 2014), biterm topic model, which works well for modeling topics in short text (Yan, Guo, Lan, & Cheng, 2013), and BERTopic, a novel method that uses class-based TF-IDF (Grootendorst, 2022). The app meticulously inspects each column within the pandas dataframe, specifically targeting those in object form, and discards any irrelevant data. This functionality proves invaluable in streamlining user analysis, particularly when handling custom files. By default, the app incorporates word stop removal and lemmatization, enhancing the usability of the data. Moreover, users have the flexibility to opt for additional enhancements (such as eliminating copyright statements, removing punctuation, and specifying stopwords for exclusion) in addition to choosing their choice of topic modeling parameters, thereby optimizing the analysis results.

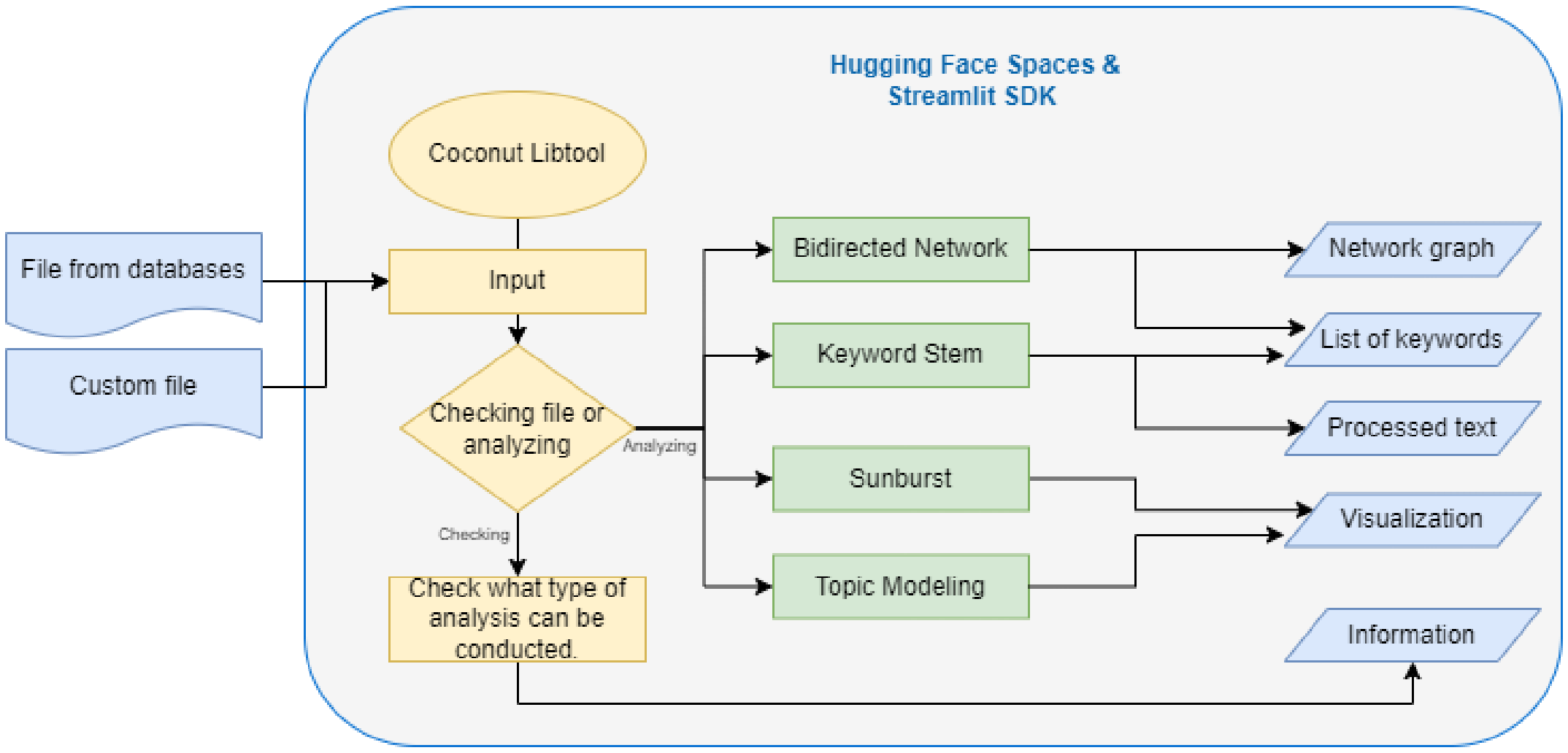

**Figure 1. Coconut Libtool App Workflow**

### Text Visualization

In Coconut Libtool, we used a 'bidirectional network' based on association rules (Santosa, 2023) and 'sunburst' (2024) for text visualization. In bidirectional network analysis, the created rules can be visualized in the form of a network graph where the included nodes can be selected according to preference, instead of having to include all nodes in the visualization. In the case of custom files, proper node separation is crucial for accurate detection. Therefore, each node must be delimited by a semicolon to ensure clear distinction and effective parsing between nodes.

Sunburst visualization helps to visualize the hierarchies between the documents and citations using Plotly (2024). It is divided into 3 layers, where the innermost layer represents the 'document type', the middle layer represents the 'source title', and the outermost layer represents the 'publication year'. The size of the slices is determined by the number of documents owned, while the color of the slice is determined by the citations. In the outer layer, citations are counted as much as it is owned, but in its inner layer, it is the weighted average of its outer layer.

### Deployment

The app was made using the python's streamlit (2024) package and was deployed using Hugging Face spaces. Coconut Libtool can be run without requiring installation or downloading the application on the user's side and can be accessed through the link Coconut Libtool, both on a computer and a mobile phone.

## USER INTERFACE FEATURES AND FUNCTIONS

Coconut Libtool consists of 2 main menus, namely (i) 'File Checker' to check what kind of analysis can be done with the user's data and the (ii) 'Analysis' menu which consists of 4 features. Generally, users can upload files by drag and drop method or by selecting files through browsing files.

In the 'Keywords Stem' feature, users can either choose (a)'lemmatization' (set as default) or (b)'stemming'. The app will automatically detect the keyword column for each bibliometric database. The output of this feature results in two options – (i) 'Result' which displays the processed data; and (ii) 'List of Keywords' which provides the file containing a list of past and modified keywords.

The ‘Topic Modeling’ feature offers a range of options, including three distinct algorithms: pyLDA, Biterm, and BERTopic. Users also have the flexibility to specify which columns to analyze from the uploaded data file. Coconut Libtool meticulously examines each relevant column for analysis, empowering users to select the column they prefer for topic modeling through the ‘choose column’ option. The primary task for users is to ascertain the optimal number of topics and, if needed, fine-tune certain advanced parameters accessible within the advanced settings. In pyLDA, users can adjust the ‘random state’ and ‘chunk size’, while in Biterm they can choose the ‘random state seed’ and ‘iteration number.’ More complex options are available for BERTopic, where users can adjust ‘top_n_words’, ‘n_components’, ‘random_state’, and ‘n_neighbors’. Additionally, users can select embedding models of their choice from these three options: (i) all-MiniLM-L6-v2, (ii) en_core_web_md, and (iii) paraphrase-multilingual-MiniLM-L12-v2. The topic modeling feature generates interactive visualizations based on the outcomes provided by each algorithm python library. (Figure 2).

Preprocessing in the ‘bidirected network’ feature is similar to the ‘Keywords Stem’ where users can set the values for ‘support’, ‘confidence’, and ‘maximum length for the generated itemsets’ to filter the number of nodes to be displayed. The filtered data then appears in the form of a table, where the selected nodes can be viewed. Users can save the visualization by right-clicking and selecting ‘save image’ option.

‘Sunburst’ is the last feature of the app which is useful for viewing the diversity of the number of articles and citations from the sources it owns. The colors indicate the depth of citations while the number of articles is depicted in the size of the slice. Filters can be used based on the time range available in the year column of the uploaded data.

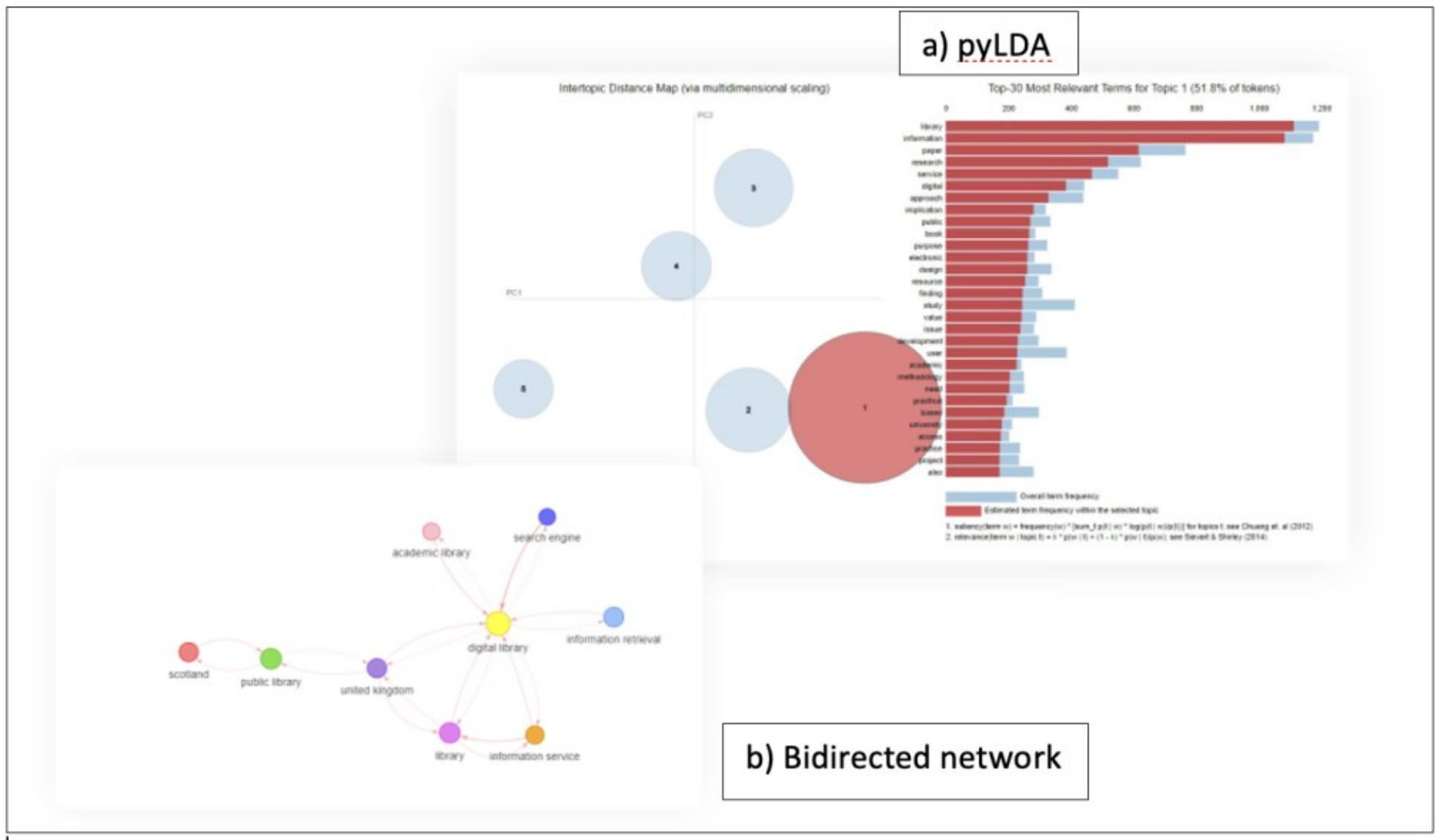

**Figure 2. Example Visualization of the Results from a) pyLDA and b) Bidirected Network**

## DISCUSSION, CONCLUSION, AND FUTURE WORK

The Coconut Libtool provides a comprehensive solution that requires no software installation and offers multiple functionalities within one interface. By supporting various analytical methods and accepting diverse types of textual data inputs, the tool promotes flexibility and customization in textual analysis. By putting people first, Coconut Libtool has significant implications for both information professionals and researchers across various disciplines by:

(i) Offering a user-friendly interface and eliminating the need for coding expertise, the Coconut Libtool makes textual analysis **more accessible** to a wider audience, including those without technical backgrounds. This democratization of access can empower individuals from diverse fields to leverage textual analysis techniques for their research or professional endeavors.

(ii) Integrating multiple analytical methods within one tool, streamlining the analysis process, and saving time and effort for users. Researchers can perform diverse analyses without switching between platforms, leading to **increased efficiency and productivity** in their work.

(iii) Fostering collaboration and interdisciplinary research efforts. Researchers from different fields can utilize the tool to analyze textual data relevant to their domains, encouraging **interdisciplinary collaboration**.

(iv) Promoting **transparency, reproducibility, accessibility, reciprocity, and responsibility** in research practices. Researchers can openly share methods and findings, fostering collaboration and knowledge sharing within their disciplinary communities.

Coconut Libtool has a few limitations that we are striving to ameliorate in our future work. We are working on: (i) expanding bibliographic data sources (e.g. Dimensions.ai and PubMed); (ii) providing more text analysis and visualization options (e.g., burst detection and scattertext); (iii) overcoming runtime constraints; and, (iv) incorporating languages beyond English. However, Coconut Libtool represents a significant advancement in the field of textual analysis, offering a user-friendly and publicly accessible solution that empowers information professionals, researchers, and learners to explore and analyze textual data more effectively.

## GENERATIVE AI USE

We confirm that we did not use generative AI tools/services to author this submission.

## AUTHOR ATTRIBUTION

FAS: conceptualization, formal analysis, software, methodology, data curation, writing-original draft

ML: conceptualization, investigation, project administration, supervision, writing-original draft, writing-review & editing

CG: methodology, writing-original draft

JSD: supervision, writing-review & editing

## REFERENCES

ACRL Research Planning and Review Committee. (2018). 2018 top trends in academic libraries: A review of the trends and issues affecting academic libraries in higher education. College & Research Libraries News, 79(6), 286. https://doi.org/10.5860/crln.79.6.286

ACRL Research Planning and Review Committee. (2020). 2020 top trends in academic libraries: A review of the trends and issues affecting academic libraries in higher education. College & Research Libraries News, 81(6), 270. https://doi.org/10.5860/crln.81.6.270

Aria, M., & Cuccurullo, C. (2017). bibliometrix: An R-tool for comprehensive science mapping analysis. Journal of Informetrics, 11(4), 959–975. https://doi.org/10.1016/j.joi.2017.08.007

Aria, M., Cuccurullo, C., D’Aniello, L., Misuraca, M., & Spano, M. (2023). TAll: A new Shiny app of Text Analysis for All. CEUR Workshop Proceedings, 3596. Venice: CEUR-WS. Retrieved from https://ceur-ws.org/Vol-3596/short2.pdf

Bird, S., Klein, Ewan, & Loper, Edward. (2009). Natural Language Processing with Python (First Edition). O’reilly.

Blei, D. M. (2012). Probabilistic topic models. Communications of the ACM, 55(4), 77–84. https://doi.org/10.1145/2133806.2133826

Fan, L., Lafia, S., Li, L., Yang, F., & Hemphill, L. (2023). DataChat: Prototyping a Conversational Agent for Dataset Search and Visualization. Proceedings of the Association for Information Science and Technology, 60(1), 586–591. https://doi.org/10.1002/pra2.820

Grootendorst, M. (2022, March 11). BERTopic: Neural topic modeling with a class-based TF-IDF procedure. arXiv. https://doi.org/10.48550/arXiv.2203.05794

Heldens, S., Sclocco, A., Dreuning, H., Van Werkhoven, B., Hijma, P., Maassen, J., & Van Nieuwpoort, R. V. (2022). litstudy: A Python package for literature reviews. SoftwareX, 20, 101207. https://doi.org/10.1016/j.softx.2022.101207

Jiang, J., & Srinivasan, K. (2023). MoreThanSentiments: A text analysis package. Software Impacts, 15, 100456. https://doi.org/10.1016/j.simpa.2022.100456

Lamba, M., Madhusudhan, M. (2022). Tools and Techniques for Text Mining and Visualization. In: Text Mining for Information Professionals. Springer, Cham. https://doi.org/10.1007/978-3-030-85085-2_10

McLevey, J., & McIlroy-Young, R. (2017). Introducing metaknowledge: Software for computational research in information science, network analysis, and science of science. Journal of Informetrics, 11(1), 176–197. https://doi.org/10.1016/j.joi.2016.12.005

Milev, P. (2023). The Role of Data Visualization in Enhancing Textual Analysis. Trakia Journal of Sciences, 21, 248–253. https://doi.org/10.15547/tjs.2023.s.01.042

Müller, M. (2023). PyblioNet – Software for the creation, visualization and analysis of bibliometric networks. SoftwareX, 24, 101565. https://doi.org/10.1016/j.softx.2023.101565

Pereira, V., Basilio, M. P., & Santos, C. H. T. (2023, April 27). pyBibX -- A Python Library for Bibliometric and Scientometric Analysis Powered with Artificial Intelligence Tools. arXiv. Retrieved from http://arxiv.org/abs/2304.14516

Plotly. (2024). Retrieved from (https://plotly.com/python/).

Rose, M. E., & Kitchin, J. R. (2019). pybliometrics: Scriptable bibliometrics using a Python interface to Scopus. SoftwareX, 10, 100263. https://doi.org/10.1016/j.softx.2019.100263

Santosa, F. A. (2023). Adding Perspective to the Bibliometric Mapping Using Bidirected Graph. Open Information Science, 7(1), 20220152. https://doi.org/10.1515/opis-2022-0152

Santosa, F. A., Lamba, M., & George, C. (2023). Coconut Library Tool: A Web-Based Application for Advanced Textual Analysis for Librarians. Retrieved from https://hdl.handle.net/2142/122277

Seadle, M. S. (2016). Managing and mining historical research data. Library Hi Tech, 34(1), 172–179. https://doi.org/10.1108/LHT-09-2015-0086

Sievert, C., & Shirley, K. (2014). LDAvis: A method for visualizing and interpreting topics. Proceedings of the Workshop on Interactive Language Learning, Visualization, and Interfaces, 63–70. Baltimore, Maryland, USA: Association for Computational Linguistics. https://doi.org/10.3115/v1/W14-3110

Streamlit. (2024). Retrieved from (https://streamlit.io/).

Sunburst. (2024). Retrieved from (https://plotly.com/python/sunburst-charts/).

Thakur, K., & Kumar, V. (2021). Application of Text Mining Techniques on Scholarly Research Articles: Methods and Tools. New Review of Academic Librarianship, 1–25. https://doi.org/10.1080/13614533.2021.1918190

Velasquez, J. D. (2023). TechMiner: Analysis of bibliographic datasets using Python. SoftwareX, 23, 101457. https://doi.org/10.1016/j.softx.2023.101457

Yan, X., Guo, J., Lan, Y., & Cheng, X. (2013). A Biterm Topic Model for Short Texts. Proceedings of the 22nd International Conference on World Wide Web, 1445–1456. Brazil: Association for Computing Machinery. https://doi.org/10.1145/2488388.2488514